\documentclass[graybox]{svmult}

\usepackage{mathptmx}       % selects Times Roman as basic font
\usepackage{helvet}         % selects Helvetica as sans-serif font
\usepackage{courier}        % selects Courier as typewriter font
\usepackage{type1cm}        % activate if the above 3 fonts are
\usepackage{makeidx}         % allows index generation
\usepackage{graphicx}        % standard LaTeX graphics tool
\usepackage{multicol}        % used for the two-column index
\usepackage[bottom]{footmisc}% places footnotes at page bottom

\usepackage{rotating}

\makeindex             % used for the subject index
\usepackage{natbib}

\begin{document}

\bibliographystyle{spbasic}

\title*{Stellar Streams and Clouds in the Galactic Halo}
% Use \titlerunning{Short Title} for an abbreviated version of
% your contribution title if the original one is too long
\author{Carl J. Grillmair and Jeffrey L. Carlin}
% Use \authorrunning{Short Title} for an abbreviated version of
% your contribution title if the original one is too long
\institute{Spitzer Science Center, Mail Stop 220-6, 1200 E. California Blvd., Pasadena, CA 91125, U.S.A.; \\
Rensselaer Polytechnic Institute, 110 8th Street, Troy, NY 12180, U.S.A. \\
%\email{carl@ipac.caltech.edu; carlij@rpi.edu}
}
%
% Use the package "url.sty" to avoid
% problems with special characters
% used in your e-mail or web address
%

\motto{Chapter~4 from the volume ``Tidal Streams in the Local Group and Beyond: Observations and Implications''; ed. Newberg, H.~J., \& Carlin, J.~L.\ 2016, Springer International Publishing, Astrophysics and Space Science Library, 420\\ ISBN 978-3-319-19335-9; DOI 10.1007/978-3-319-19336-6\_1}

\maketitle

\abstract{Recent years have seen the discovery of an ever growing
  number of stellar debris streams and clouds. These structures are
  typically detected as extended and often curvilinear overdensities of
  metal-poor stars that stand out from the foreground disk
  population. The streams typically stretch tens of degrees or more
  across the sky, even encircling the Galaxy, and range in
  heliocentric distance from 3 to 100 kpc.  This chapter summarizes
  the techniques used for finding such streams and provides tables
  giving positions, distances, velocities, and metallicities, where
  available, for all major streams and clouds that have been detected
  as of January 2015.  Sky maps of the streams are also provided.
  Properties of individual tidal debris structures are discussed.}

\section{Observational techniques}
\label{sec:1}

With the advent of wide-area, digital sky surveys (primarily the Sloan
Digital Sky Survey\index{Sloan Digital Sky Survey (SDSS)}, SDSS, as well as the 2MASS\index{Two Micron All Sky Survey (2MASS)} and WISE\index{Wide-Field Infrared Survey Explorer (WISE)} surveys), we
have seen a remarkable period of discovery in the search for Galactic
substructures. From the initial detection of short tidal tails
emanating from the globular cluster Palomar 5\index{Palomar 5} in the SDSS
commissioning data (Odenkirchen et al. 2001), to the mapping of
M-giants\index{M giants} in the Sagittarius stream\index{Sagittarius (Sgr) stream} completely around the sky using
2MASS (Majewski et al. 2003), our knowledge of coherent structures in
the halo of our Galaxy has advanced spectacularly. While there have been many
previous attempts to map large regions of sky in search of tidal
streams and substructures, it appears in hindsight that these efforts were
largely defeated by insufficient or non-contiguous sky coverage,
variable depth or completeness, or insufficiently uniform calibration.

The streams and clouds described here were all initially detected
using photometric techniques\index{streams!photometric searches}. These techniques rely on the separation
in color-magnitude space of relatively low-metallicity\index{metallicities} halo stars from
the far larger population of nearby foreground stars. Early efforts to
detect tidal debris streams applied these techniques to photographic
surveys of regions surrounding globular clusters and dwarf galaxies
\citep{grillmair95, irwin95, leon00}. The same techniques are now
being applied to deeper, wider, and better calibrated large-scale
digital surveys (e.g., SDSS, 2MASS, WISE, PAndAS\index{Pan-Andromeda Archaeological Survey (PAndAS)}, and Pan-STARRS\index{Panoramic Survey Telescope and Rapid Response System (Pan-STARRS)}).
Other halo substructures have been detected in radial velocity\index{radial velocities} surveys
as cold (low velocity dispersion), coherent subsets of local halo
stars; these are described in Chapter 5.

Early efforts at separating halo populations from foreground stars
relied on fairly simple color and magnitude cuts designed to select
blue main sequence turn-off stars. This technique was used with some
success to find tidal streams emanating from globular clusters
(Grillmair et al. 1995, Leon et al. 2000, Odenkirchen et
al. 2001). Rockosi et al. (2002) subsequently demonstrated the utility
of a matched filter\index{matched filter} in making better use of the available information
and pushing detection limits to considerably lower surface
densities. By assigning weights or likelihoods to individual stars
based on their photometric uncertainties and the color-magnitude
distribution of nearby field stars, the matched filter yields the optimal
contrast between different stellar populations. Grillmair and Dionatos
(2006a, 2006b) adopted this technique to extend the Pal 5
tidal stream and to discover the tenuous and very cold stream
GD-1\index{GD-1}. More sophisticated techniques are combining photometric
selection with radial velocity surveys (e.g., Newberg et al. 2009) to
push detection limits to still lower surface densities.

Several stream candidates have been detected purely photometrically as
long overdensities in the SDSS and WISE surveys and await confirmation
via velocity surveys. Given the low surface densities of known
streams, and the fact that stars can only be assigned probabilities of
being associated with either the field or the stream, it can be a
fairly expensive proposition to obtain velocity information. Telescope
time allocation committees are loath to grant large blocks of time
when the anticipated yield might be as little as 10\% of the targets
surveyed. Confirmation of some of the most tenuous streams may
consequently have to wait for results of massively multiplexed
spectroscopic surveys such as LAMOST\index{Large Sky Area Multi-Object Fiber Spectroscopic Telescope (LAMOST)} \citep{czc+12, dnl+12, lzz+12,
  zzc+12} and DESI\index{Dark Energy Spectroscopic Instrument (DESI)} (Dark Energy Spectroscopic Instrument; formerly
known as BigBOSS).

Even while we wait for spectroscopic confirmation, more stream
candidates are actively being sought. While tightly constraining the
shape\index{Galactic potential!shape of} and extent of the Galactic potential with tidal streams will
ultimately require full, six-dimensional phase space information
(e.g. Willett et al. 2009, Koposov et al. 2010, Law \& Majewski 2010),
simply finding more streams will enable us to better understand the
accretion history of the Galaxy. The distances and orientations of the
streams give us some idea of their orbits and the likely distribution
of their progenitors. And if we can properly quantify the current
detection biases, we will also be in a better position to infer the
total population of such streams and their progenitors.

\section{Currently known stellar debris streams and clouds in the Galactic halo}
\label{sec:2}

Here we present tables that give some basic information for all of the
currently known streams\index{streams} and clouds.  We define streams as
overdensities of stars that are significantly longer than they are
wide.  Where radial velocities\index{radial velocities} have been measured, they have a narrow
line-of-sight velocity dispersion\index{velocity dispersion} that can vary with position along
the stream.  Narrower, lower velocity dispersion streams are thought
to result from the tidal disruption of globular clusters\index{globular clusters}, while wider,
hotter streams are thought to result from the disruption of dwarf
galaxies\index{dwarf galaxies}.  Clouds\index{clouds}, on the other hand, are spatially very large
overdensities of halo stars whose origins are currently not understood. These
clouds may be the tidal debris from dwarf galaxies that has piled up
near the apogalactica of highly eccentric orbits. There is also 
evidence that at least one cloud is the disrupted remains of the core
of the dwarf galaxy.

Tables 1 and 2 are primarily intended to 1) aid non-specialists
in identifying unexpected featues in their data, and 2) provide
interested researchers with some basic parameters and references to
more detailed works. The tables are neither intended nor usable for
comparative studies.

Table 1 summarizes the basic properties of all currently known halo
debris streams in order of discovery. We do not include globular or
open clusters with power-law profiles extending beyond the nominal
King tidal radii. \citet{grillmair95} and \citet{leon00} found such extensions
around most of the globular clusters in their respective
surveys. Given both these results and theoretical expectations, it is
probably fair to say that it would be surprising to find a globular
cluster that, if examined sufficiently deeply, did {\it not} show
evidence of tidal stripping. We therefore list only streams and
structures with a projected extent of at least a few degrees on the
sky.  We report all tidal streams discovered in the literature,
including some that remain controversial.  For example, the Monoceros
Ring\index{Monoceros Ring} could actually be a feature of the Milky Way's stellar disk (see
Chapter 3).  The Virgo Stellar Stream\index{Virgo Stellar Stream (VSS)} (VSS) and the Virgo Overdensity\index{Virgo Overdensity (VOD)}
(VOD) could be the same structure, or several structures that overlap
in configuration space.  As several deep, wide-area surveys are in
progress in both hemispheres, and as the search for stellar
substructures remains quite active, we expect this table to
become incomplete fairly quickly.

Column descriptions for Tables 1 and 2 are as follows:

%
%\noindent Column Descriptions:\\

{\it Designation:} This column gives the common name by which the
particular feature is known in the literature. There exists as yet no
convention on how streams should be named\index{naming of streams} and a delightful anarchy has
ensued. In cases where the progenitor is known, the feature is quite
naturally named after the progenitor (e.g., Sagittarius stream, Pal 5,
NGC 5466).  Some researchers have named streams after the
constellations or regions of the sky in which they were first found,
or in which they currently appear strongest (e.g., Monoceros, Cetus
Polar Stream, Anticenter Stream, Triangulum/Pisces, Ophiuchus). Still
others are named for the surveys in which they were first detected
(e.g., ATLAS stream, PAndAS MW stream), after some discovery
characteristic (Orphan, EBS), or after rivers in Greek mythology
(Acheron, Lethe, Cocytos, Styx, Alpheus, Hermus, \& Hyllus).

{\it Progenitor:} This column provides the name of the progenitor\index{progenitors of streams}, if
known or suspected. Where there is a question mark, readers are
referred to the references for the source of the uncertainty. If the
progenitor is listed as dG? (dwarf galaxy) or GC? (globular cluster),
the actual progenitor is unknown, but its likely nature is conjectured
based on the strength or morphology of the stream. Extensive, broad
(FWHM $> 500$ pc), or hot streams ($\sigma_v \ge 10$ km s$^{-1}$)
presumably arose from more massive systems such as dwarf galaxies,
whereas narrower (FWHM $< 200$ pc), less populous, and colder streams
($\sigma_v \le 5 $km s$^{-1}$) most likely arose from globular
clusters.\index{morphologies of debris structures!stream-like vs. cloud-like} This latter conclusion is based primarily on the similarity
in the cross-sectional widths of such streams ($\sim 100$ pc) to those
of known globular cluster streams (Pal 5, NGC 5466). On the other
hand, \citet{carlberg2009} has shown that streams should be heated by
encounters with other Galactic constituents, and that initially narrow
streams should become hotter and broader with time. Our division between
globular cluster and dwarf galaxy streams must therefore be regarded
with some skepticism until more observational evidence can be brought
to bear (e.g. progenitor identifications, $\alpha$-element ratios,
etc).

{\it Known Extent:} To help researchers to identify structures they
may come across in the course of their work, we provide the maximum
extent of the streams on the celestial sphere, as determined either in
the discovery paper or in subsequent investigations. While the
positions and trajectories of some features are shown in Figures 1
through 6, readers are generally referred to the discovery or
follow-up papers for more detailed maps of the streams. In some cases
the streams evidently extend beyond the discovery survey regions. For
these features we simply provide the R.A. and Dec limits of the
survey used.

{\it Distance:} This column provides the range of heliocentric
distances for different portions of streams. Where different distances
are estimated by different authors, we provide only the most recent
estimates. These distance estimates will undoubtedly be refined as
deeper surveys are carried out, spectroscopic metallicities are
obtained, proper motions and Galactic parallaxes are measured, or RR
Lyrae can be physically associated with streams. As distances can vary
greatly from one portion of a given stream to another, readers are
referred to discovery and follow-up papers for estimates of
distance with position.

{\it V$_{hel}$:} If radial velocities\index{radial velocities} have been measured for one or
more portions of a stream, the range of velocities is given here. This
range can be quite large (e.g., Sagittarius or GD-1), where the length
of the stream, combined with the Sun's motion, can produce very
substantial gradients over the length of the stream. Once again,
readers are referred to the references for more specific information.

{\it [Fe/H]:} Metallicity\index{metallicities} estimates can be based on the
color-magnitude locus of stream stars, on the metallicity of the
matched filter that yields the highest signal-to-noise ratio, or on
spectroscopy of individual stars. Estimates based solely on photometry
can have fairly large uncertainties and are consequently flagged with
question marks.

{\it Selected References:} We provide a selected set of references for
each stream. These include the discovery paper(s) and subsequent works
that demonstrably extend the stream, provide additional spectroscopic
or proper motion measurements, examine stream morphology, RR Lyrae
content, model the orbits of the streams, or use the streams as probes
of the Galactic potential.

Clouds are listed in Table 2, and are distinguished from streams by
their more diffuse, non-localized nature. These features are also
likely to be the result of tidal stripping or disruption, but
determining their nature and origin may have to await future radial
velocity and proper motion surveys.

\begin{table}
%\begin{longtable}{llccccl}
%\caption{Currently Known Halo Streams}
\caption{Currently Known Halo Streams\index{known streams!table of}}
\label{tab:1}       % Give a unique label
\resizebox{\textwidth}{!}{
\begin{tabular}{llccccl}
\hline\noalign{\smallskip}

\small

Designation & Progenitor\index{known streams!progenitors} & Known Extent\index{known streams!spatial extent} & Distance\index{known streams!distances} & V$_{hel}$\index{known streams!velocities} & [Fe/H]\index{known streams!metallicities} & Selected References \\
& & & kpc & km s$^{-1}$ & & \\
\noalign{\smallskip}\svhline\noalign{\smallskip}
Sagittarius\index{Sagittarius (Sgr) stream}   & Sagittarius & Circum-sky & 7-100 & (-200, +200) & (-1.15, -0.4) & Ibata et al. 1994, Mateo et
al. 1998, \\
Stream & dSph & & & & & Alard 1996, Totten \& Irwin 1998, \\
& & & & & & Ibata et al. 2001a, b,  Majewski et al. 2003,  \\
& & & & & & Martinez-Delgado et al. 2004, \\
& & & & & & Vivas et al. 2005, Belokurov et al. 2006b,  \\
& & & & & & Fellhauer et al. 2006, Bellazzini et al. 2006b, \\
& & & & & & Chou et al. 2007, Law et al. 2009, \\
& & & & & & Law \& Majewski 2010, Keller et al. 2010\\
& & & & & & Carlin et al. 2012a, Koposov et al. 2012\\
\\

Virgo Stellar\index{Virgo Stellar Stream (VSS)} & NGC 2419?\index{NGC 2419} & $180^{\circ} <$ R.A. $< 195^{\circ}$ & 19.6 & 128  & -1.78 & Vivas et al. 2001, Duffau et
al., 2006, \\
Stream & & $-4^{\circ} < \delta < 0^{\circ}$  & & & & Newberg et al. 2007, Duffau et al. 2014  \\
\\

Palomar 5 \index{Palomar 5}    & Palomar 5  & $225^{\circ} < $ R.A. $< 250^{\circ}$ & 23 & -55  & -1.43 & Odenkirchen et al. 2001,
2003, 2009, \\
& & $-3^{\circ} < \delta < 8.5^{\circ}$  & & & & Rockosi et al. 2002,  \\
& & & & & & Grillmair \& Dionatos 2006a, \\
& & & & & & Carlberg, Grillmair, \& Hetherington 2012 \\
\\

Monoceros Ring \index{Monoceros Ring} & dG? & $108^{\circ} <$ R.A. $< 125^{\circ}$ & $\approx 10.5$ & $\approx 100$ & -0.8 & Newberg et al. 2002,
Yanny et al. 2003, \\
& & $-3^{\circ} < \delta < -41^{\circ}$ & & & & Ibata et al. 2003, Rocha-Pinto et al. 2003, \\
& & & & & & Penarrubia et al. 2005, Li et al. 2012 \\
& & & & & & Slater et al. 2014 \\
\\

NGC 5466 \index{NGC 5466} & NGC 5466 & $182^{\circ} < $ R.A. $ < 224^{\circ}$ & 17 & 108 & -2.2  &  Belokurov et al. 2006a,\\
& & $21^{\circ} < \delta < 42^{\circ}$  & & & &  Grillmair \& Johnson 2006, \\
& & & & & & Fellhauer et al. 2007b, Lux et al. 2012 \\
\\

Orphan Stream\index{Orphan Stream} & dG? & $143^{\circ} < $ R.A. $< 165^{\circ}$ & 20-55 & (95,240) & -2.1 & Grillmair 2006a, Belokurov et al. 2007a \\
& & $-17^{\circ} < \delta < +48^{\circ}$ & & & & Fellhauer et al. 2007a, Sales et al. 2008, \\
& & & & & & Newberg et al. 2010, Sesar et al. 2013 \\
& & & & & & Casey et al. 2013 \\
\\

GD-1     \index{GD-1}     & GC? & $134^{\circ} < $ R.A. $ < 218^{\circ}$ & 7 - 10 & (-200,+100) &  -2.1 & Grillmair \& Dionatos
2006b, \\
& & $14^{\circ} < \delta < 58^{\circ}$ & & & & Willett et al. 2009, \\
& & & & & & Koposov, Rix, \& Hogg 2010, \\
& & & & & &  Carlberg \& Grillmair 2013 \\
\\

AntiCenter Stream\index{Anticenter Stream (ACS)}& dG? & $121^{\circ} < $ R.A. $< 130^{\circ}$ & $\approx 8$ & (50, 90)  & -0.96 & Grillmair 2006b, \\
& & $-3^{\circ} < +63^{\circ}$ & & & & Grillmair, Carlin, \& Majewski 2008, \\
& & & & & & Li et al. 2012\\
\\

EBS     \index{Eastern Banded Structure (EBS)}      & GC? & $132^{\circ} < $ R.A. $ < 137^{\circ}$ & 10 & (71, 85)  & -1.8 & Grillmair 2006b, Grillmair 2011,\\
& & $-3^{\circ} < \delta <  16^{\circ}$ & & & & Li et al. 2012\\
\\

Acheron  \index{Acheron}     & GC? & $ 232^{\circ} < $ R.A. $ < 258^{\circ}$ & 3.5 - 3.8 & & -1.7? & Grillmair 2009 \\
& & $3^{\circ} < \delta < 20^{\circ}$  & & & &\\
\\

Cocytos   \index{Cocytos}    & GC? & $ 197^{\circ} < $ R.A. $ < 257^{\circ}$ &  11 & & -1.7? & Grillmair 2009 \\
& & $8^{\circ} < \delta < 30^{\circ}$ & & & & \\
\\

Lethe    \index{Lethe}     & GC? & $ 176^{\circ} < $ R.A. $ < 252^{\circ}$ & 13  & & -1.7? &  Grillmair 2009 \\
& & $22^{\circ} < \delta < 37^{\circ}$ & & & & \\
\\

Styx     \index{Styx}     & Bootes III?\index{Bo{\"o}tes III}           & $201^{\circ} < $ R.A. $ < 250^{\circ}$, & 45  & & -2.2? & Grillmair 2009 \\
& & $21^{\circ} < \delta < 31^{\circ}$  & & & & \\
\\

Cetus Polar \index{Cetus Polar Stream} & NGC 5824?\index{NGC 5824}  & $19^{\circ} < $ R.A. $ < 37^{\circ}$ & 24 - 36 & (-200, -160) & -2.1 & Newberg, Yanny, \& Willett 2009, \\
Stream & & $-11^{\circ} < \delta < +39^{\circ}$  & & & & Yam et al. 2013 \\
\\

Pisces/ \index{Pisces/Triangulum Stream} & GC? & $21^{\circ} < $ R.A. $< 24^{\circ}$ & 35  & 120  & -2.2
& Bonaca et al. 2012b, Martin et al. 2013, \\
Triangulum & & $23^{\circ} < \delta < 40^{\circ}$ & & & & Martin et al. 2014\\
\\

Alpheus   \index{Alpheus}    & NGC 288?\index{NGC 288}  & $ 22^{\circ} < $ R.A. $< 28^{\circ}$ & 1.6 - 2.0  & & -1.0?     & Grillmair et al. 2013\\
& & $ -69^{\circ} < \delta < 45^{\circ}$ & & & & \\
\\

ATLAS stream\index{ATLAS stream}  & Pyxis?\index{Pyxis}     & $18^{\circ} < $ R.A. $< 30^{\circ}$ & 20  & & -1.4?   & Koposov et al. 2014 \\
& & $ -32^{\circ} < \delta < -25^{\circ}$ & & & & \\
\\

PAndAS MW  \index{PAndAS MW Stream}&  dG? & $ 0^{\circ} < $ R.A. $< 22^{\circ}$  & 17 & 127 & -1.5? & Martin et al. 2014\\
Stream & & $40^{\circ} < \delta < 48^{\circ} $ \\
%\\
\\

Hermus   \index{Hermus}     & GC? & $241^{\circ} < $ R.A. $ < 254^{\circ}$ & 18.5  & & -2.3? & Grillmair 2014 \\
& & $5^{\circ} < \delta < 50^{\circ}$ & & & & \\
\\

Hyllus   \index{Hyllus}     & GC? & $245^{\circ} < $ R.A. $ < 249^{\circ}$ & 20  &  & -2.3? & Grillmair 2014 \\
& & $ 11^{\circ} < \delta < 34^{\circ}$  & & & & \\
\\

Ophiuchus stream\index{Ophiuchus Stream} & GC & $241^{\circ} < $ R.A. $ < 243^{\circ}$ & 8 - 9.5 & 290  & -1.95 & Bernard et al. 2014, Sesar et al. 2015\\
& & $-7.2^{\circ} < \delta < 6.7^{\circ}$ & & & & \\

%\hline
\noalign{\smallskip}\hline\noalign{\smallskip}
\end{tabular}
}
\end{table}
%\end{longtable}

\begin{table}[!t]
%\begin{longtable}{llccccl}
%\caption{Currently Known Halo Clouds}
\caption{Currently Known Halo Clouds\index{known debris clouds!table of}}
\label{tab:2}       % Give a unique label
\resizebox{\textwidth}{!}{
\begin{tabular}{llccccl}
\hline\noalign{\smallskip}

\small

Designation & Progenitor\index{known debris clouds!progenitors} & Known Extent\index{known debris clouds!spatial extent} & Distance\index{known debris clouds!distances} & V$_{hel}$\index{known debris clouds!velocities} & [Fe/H]\index{known debris clouds!metallicities} & Selected References \\
%Designation & Progenitor & Known Extent & Distance & V$_{hel}$ & [Fe/H] & Selected References \\
& & & kpc & km s$^{-1}$ & & \\
\noalign{\smallskip}\svhline\noalign{\smallskip}
%\hline

Triangulum- \index{TriAnd1}& dG? & $-10^{\circ} < $ R.A. $ < 30^{\circ}$ & 20 & (-200, -50) &     -0.6        & Majewski et al. 2004, Rocha-Pinto et al. 2004, \\
Andromeda (TriAnd1) & & $ 20^{\circ} < \delta < 45^{\circ} $ & & &  & Martin et al. 2007, Chou et al. 2011 \\
& & & & & & Deason et al. 2014, Sheffield et al. 2014\\
\\

Hercules-Aquila \index{Hercules-Aquila Cloud}& dG? & $0^\circ < l < 80^\circ$& 10-25  & (-130, -120) & (-2.2, -1.4) & Belokurov et al. 2007b, Watkins et al. 2009, \\
& & $-50^\circ < b < 50^\circ$ & & & & Sesar et al. 2010a, Simion et al. 2014\\
\\

Virgo\index{Virgo Overdensity (VOD)} & dG? & see map in Figure 3 & $\sim$6-20  & (200, 360) & (-2.0, -1.0) & Vivas et al. 2001, Newberg et al. 2002, \\
Overdensity & &  & & & & Juric et al. 2008, Bonaca et al. 2012a, \\
& & & & & & Carlin et al. 2012b, Duffau et al. 2014\\
\\

Triangulum \index{TriAnd2} & dG?     & $ 3^{\circ} < $ R.A. $< 23^{\circ}$ & 28 & (-200, -50) & -0.6        & Martin et al. 2007, Sheffield et al. 2014 \\
-Andromeda 2 (TriAnd2) & & $ 28^{\circ} < \delta < 42^{\circ} $\\
\\

Pisces Overdensity\index{Pisces Overdensity} & dG?     & $350^{\circ} < $ R.A. $ < 360^{\circ}$
& 80-100 & -75 &  -1.5  & Sesar et al. 2007, Watkins et al. 2009,\\
&& $-1.3^{\circ} < \delta < +1.3^{\circ} $&&& &  Kollmeier et al. 2009, Sesar et al. 2010b,\\
&&&&& &  Sharma et al. 2010, Vivas et al. 2011\\

%Perseus Cloud & dG? & & 25 & 57 & -1.5 & Rocha-Pinto et al. 2003, 2004\\

%Argo Cloud & dG? & $ 93^{\circ} < $ R.A. $< 215^{\circ}$ & 14 & & -0.7 &
%Rocha-Pinto et al. 2006\\
%& & $-71^{\circ} < \delta < 0^{\circ}$ \\

%Canis Major   & dG?& $95^{\circ} < $ R.A. $< 120^{\circ}$ & 7.2 & 109
%& -0.5 & Martin et al 2004a,b, Bellazzini et al. 2004, 2006, \\
%& & $-45^{\circ} < \delta < -10^{\circ}$ & & & & Dinescu et al. 2005 \\

%\hline
\noalign{\smallskip}\hline\noalign{\smallskip}
\end{tabular}
}
\end{table}
%\end{longtable}

\normalsize

The locations and extent of many of the structures in Tables 1 and 2
are shown in Figures 1 and 2. The PAndAS MW, ATLAS, and Ophiuchus
streams are shown separately in Figures 3, 4, and 5.

%\clearpage 

\begin{center}
\begin{sidewaysfigure}[!ht]
%\begin{landscape}
%\begin{figure}[!t]
%\includegraphics[scale=0.5]{stream_map_north.eps}
%\includegraphics[width=1.0\textwidth]{stream_map_north.eps}
%\begin{figure}[!t]
%\centering
\includegraphics[width=1.0\textheight]{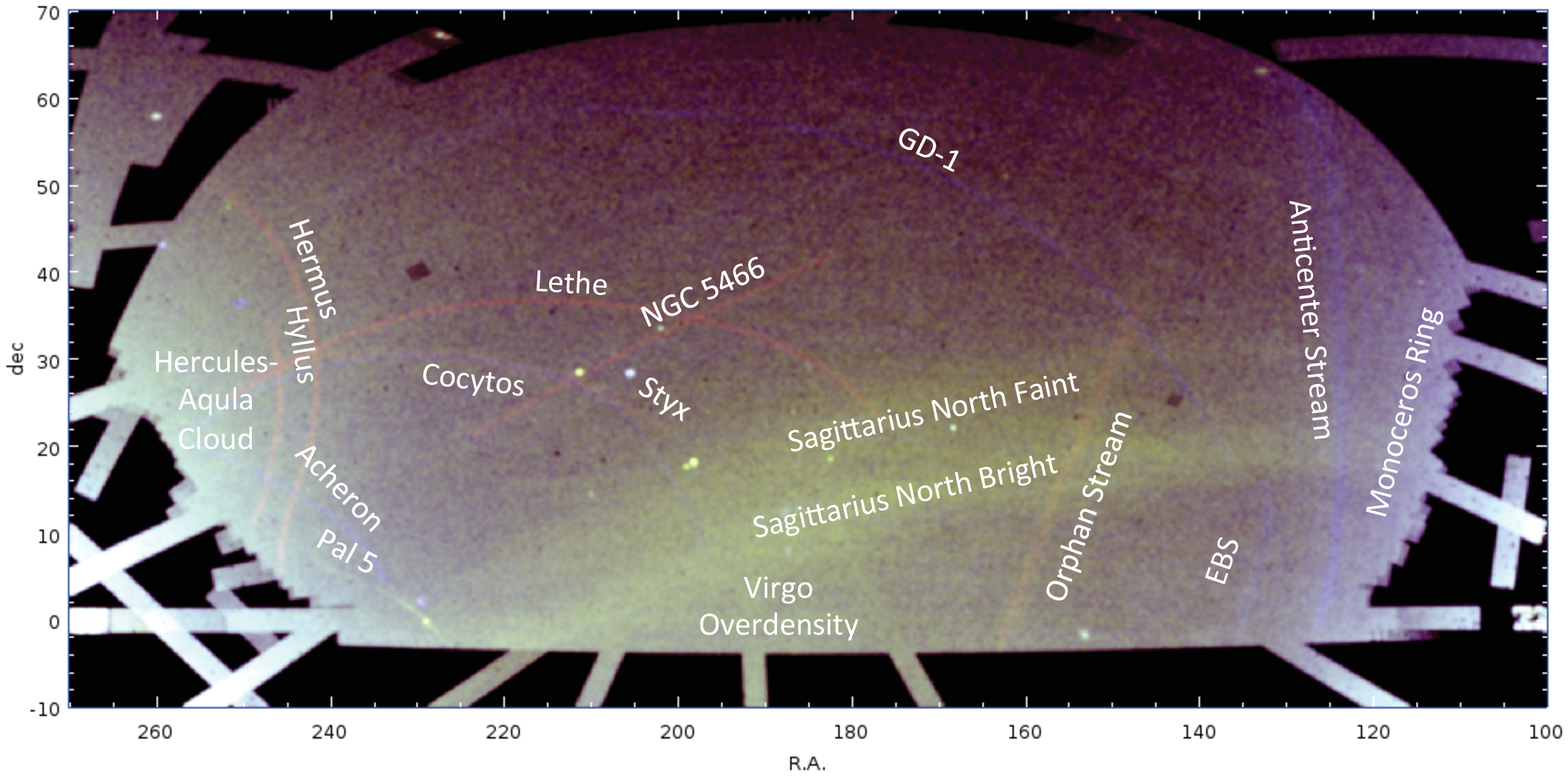}
\caption{\index{Map of known streams!northern SDSS footprint}A matched-filter surface density map of the northern footprint of
  the Sloan Digital Sky Survey\index{Sloan Digital Sky Survey (SDSS)}. The filter used here is based on the
  color-magnitude distribution of the metal poor globular cluster M
  13. The stretch is logarithmic and all but the Sagittarius streams\index{Sagittarius (Sgr) stream}
  have been enhanced using arbitrarily scaled Gaussians to make them
  visible at this stretch. Bluer colors correspond to more nearby
  stars ($\le 15$ kpc) while redder colors reflect the distribution of
  more distant stars.}
\label{fig:1}
\end{sidewaysfigure}
%\end{figure}
%\end{landscape}
\end{center}

%\clearpage 

\begin{center}
%\begin{figure}[!t]
\begin{sidewaysfigure}[!ht]
\includegraphics[width=1.0\textheight]{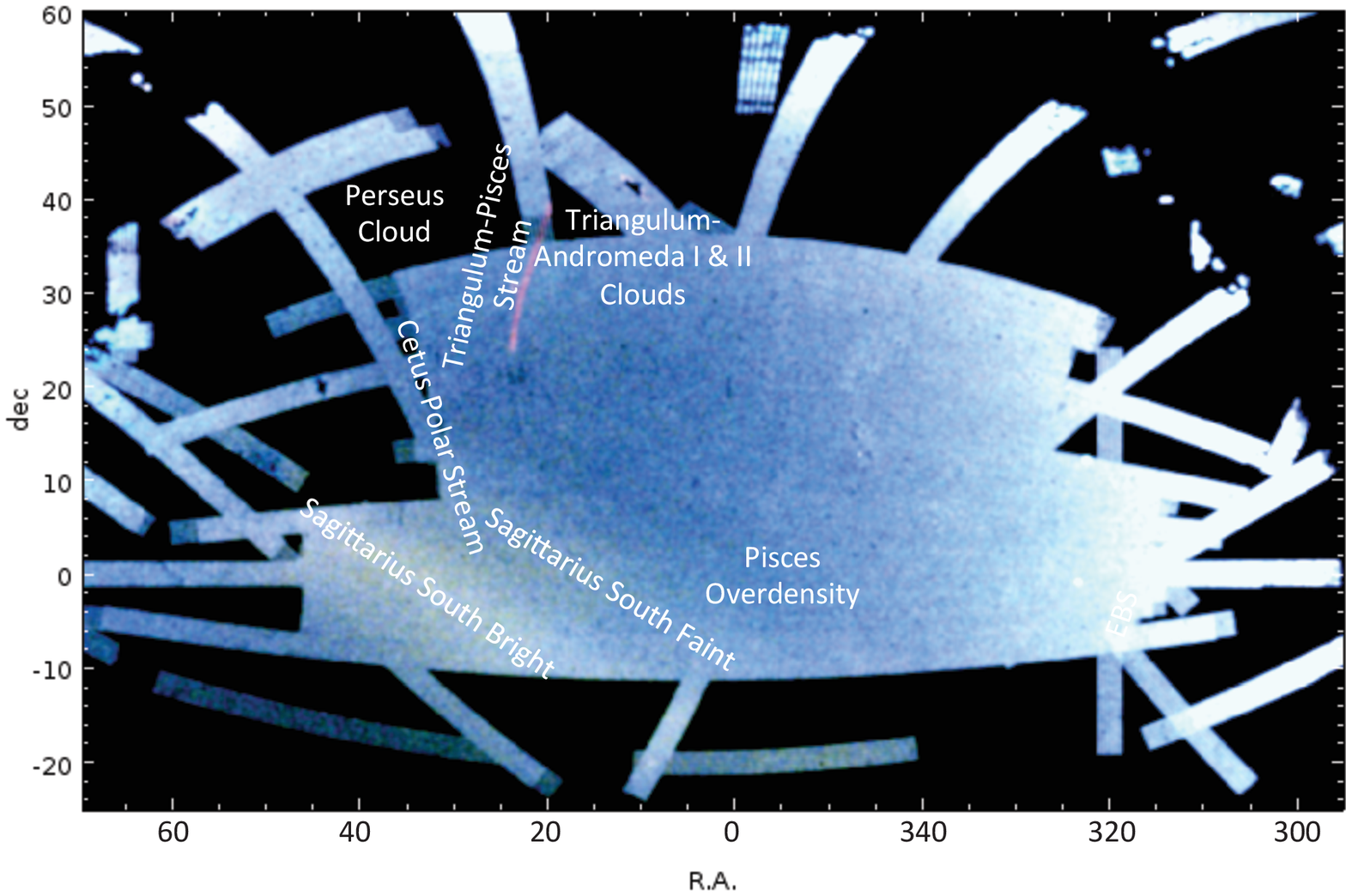}

\caption{\index{Map of known streams!southern SDSS footprint}As in Figure 1, but for the southern footprint of the SDSS\index{Sloan Digital Sky Survey (SDSS)}.}

\label{fig:2}
\end{sidewaysfigure}
\end{center}

\clearpage

In the remainder of this chapter we discuss some of the particulars of
individual streams listed in Tables~\ref{tab:1} and
\ref{tab:2}. As the Sagittarius and Monoceros/Anticenter streams
are the subject of much study and controversy, they merit
more detailed discussion, and we refer the reader to
Chapters~2 and 3 for more on these systems.

\subsection{Streams with known or likely globular cluster progenitors\index{progenitors of streams!globular clusters}}
\label{sec:GC_streams}

{\bf Pal 5:} The stream associated the globular cluster Palomar~5\index{Palomar 5}
(Pal~5) is one of the most striking examples among the known tidal
streams. The stream is clearly visible as a narrow ribbon of stars
arcing across the corner of SDSS starcount maps such as
Figure~\ref{fig:1} and the ``Field of Streams''\index{SDSS Field of Streams} from
Belokurov~et~al.~(2006b; also reproduced in Chapter~1 of this
volume). As it is clearly associated with the cluster, is a
well-defined, kinematically-cold stream, and is at a distance
($\sim23$~kpc) close enough to be studied in detail, this stream has
been the object of intense scrutiny. Indeed, as it can be studied in
such exquisite detail, the Pal~5 stream has become something of a test
case for models of tidal disruption (e.g., Dehnen~et~al.~2004;
Mastrobuono-Battisti~et~al.~2012) and evolution of the tidal tails in
the Galactic halo.  The stream is clearly defined over $>25^\circ$ on
the sky, with radial velocity\index{radial velocities} members now identified over at least
$\sim20^\circ$ of its extent (e.g., Odenkirchen~et~al.~2009;
Kuzma~et~al.~2015). As it is robustly detected above the background
stellar density, the Pal~5 stream has been used to explore gaps in
streams\index{stream gaps} and their implications for the density of dark matter subhalos\index{dark matter subhalos}
(\citealt{carlberg2012}; see also Chapter 7). As much of the stream is
still in the vicinity of Pal 5, with a rather periodic distribution of
clumps near the cluster, there remains some controversy concerning the
extent to which these gaps may be a product of epicyclic motions of
stars within the stream (K{\"u}pper et al. 2008, 2012) rather than the
result of encounters with subhalos.

{\bf NGC 5466:}\index{NGC 5466} The tidal stream associated with the globular cluster NGC~5466 has been suggested \citep{lux2012} to have strong implications for the shape of the Galactic dark matter halo\index{Galactic potential!shape of}. 
Though this stream is quite strong near the cluster
\citep{belokurov2006a}, its more distant reaches are very much more tenuous
\citep{grillmair06j}. 
The analysis by \citet{lux2012} suggested that
the trajectory of the northern tip of the stream may have important
consequences for the inferred shape of the dark matter halo. More definitive results await the identification of tracers with which to derive velocity and distance to the stream at large separations from the main body of the cluster.

{\bf GD-1:}\index{GD-1} Now traced to over 80$^{\circ}$ across the northern SDSS
footprint, the ``GD-1'' stream \citep{gd06b} is the longest of the
cold streams discovered to date. At a mean distance of $\sim9$~kpc,
the stream's narrow $\sim0.5^\circ$ width corresponds to only
$\sim70$~pc in cross section, suggesting that the GD-1 progenitor was
(or is) a globular cluster. The orbit\index{orbit fitting} of this stream has been well
established using SDSS velocity\index{radial velocities} measurements by \citet{willett09}, who
found a best-fitting orbit with pericenter of $\sim14$~kpc, apocenter
of $\sim28$~kpc, and low inclination ($i \sim 35^\circ$).
\citet{koposov10} explored the constraints that this system places on
the shape and strength of the Galactic potential\index{Galactic potential!shape of}, and were able to put
tight limits on the circular velocity at the Sun's radius (V$_c = 224
\pm 13 $ km s$^{-1}$). Constraints on the shape of the halo are
comparatively weak owing to the relatively low orbital inclination and
the strong influence of the Galactic disk. Its relatively high
surface density over a large angular swath, kinematically cold
population, presumed great age, and lack of a nearby progenitor make
GD-1 an excellent candidate to search for the perturbing effects of
dark matter subhalos (see Chapter 7). The measurement of $8\pm3$
stream gaps\index{stream gaps} over $\sim8$~kpc of GD-1's extent by \citet{carlberg2013}
suggest that the Milky Way contains some 100 dark matter subhalos\index{dark matter subhalos} with
$M > 10^6$~M$_{\odot}$ within the apocenter of the GD-1
orbit. Verifying and refining this estimate will require both deeper
photometric surveys and better radial velocity sampling of GD-1\index{GD-1}.

{\bf EBS:}\index{Eastern Banded Structure (EBS)} The Eastern Banded Structure (EBS) is a $\sim18^\circ$-long
stellar overdensity discovered in the SDSS by
\citet{grillmair06a}. Because of its proximity and similar distance
($\sim10$~kpc) to the Anticenter Stream\index{Anticenter Stream (ACS)} (ACS), the EBS was originally
thought to be associated with the ACS/Monoceros structures.
\citet{grillmair2011} and \citet{li2012} showed that EBS is distinct
both in metallicity\index{metallicities} and kinematics from the ACS. Furthermore,
\citet{grillmair2011} associated two velocity structures (ECHOS)\index{Elements of Cold HalO Substructure (ECHOS)} of
\citet{schlaufman09} with EBS and determined that its orbit is likely
to be highly eccentric, unlike that of the ACS. A $\sim2^\circ$-wide
overdensity of stars within the EBS was also noted by
\citet{grillmair2011}, who suggested that this object, dubbed Hydra~I\index{Hydra I},
may be the remains of EBS's progenitor. Deeper and wider surveys of
the EBS are currently in progress and will hopefully shed more light
on the nature and origin of this stream.

\begin{figure}[!t]
\includegraphics[width=1.0\textwidth]{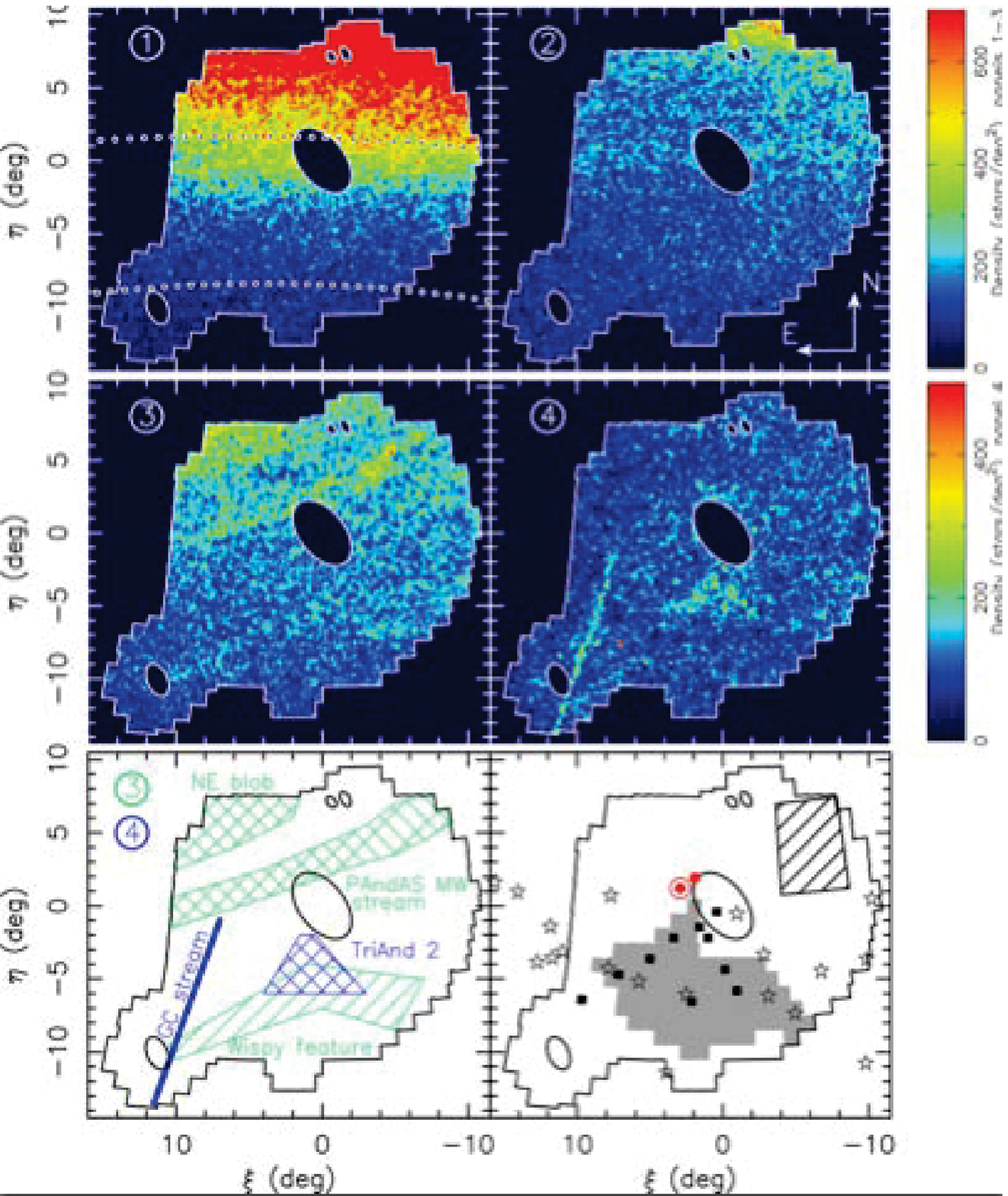}

\caption{Stellar density in the PAndAS\index{Pan-Andromeda Archaeological Survey (PAndAS)} survey in the foreground of M31 (coordinates are relative to the M31 center, with north upward and east toward the left). Panels 1-4 show smoothed stellar density maps for CMD-filtered stars at mean distances of 7, 11, 17, and 27 kpc from the Sun, respectively. The lower left panel shows the Milky Way substructures identified by \citet{martin2014} in this so-called ``PAndAS Field of Streams.''\index{PAndAS Field of Streams} [Figure~2 from \citet{martin2014}.]
}
\label{fig:pandas}
\end{figure}

{\bf Pisces/Triangulum:}\index{Pisces/Triangulum Stream} The Pisces/Triangulum Stellar Stream is a
narrow ($\sim0.2^\circ$) stream that was found as a $\sim5.5$~kpc-long
photometric overdensity by \citet{bonaca2012b}, who dubbed it the
``Triangulum Stream''\index{Triangulum Stream} based on its proximity to M33\index{M33}. It was detected
independently as a kinematically coherent feature among SDSS spectra
by \citet{martin13}, who called it the ``Pisces Stellar Stream''\index{Pisces Stellar Stream}
because it resides in Pisces. While \citet{bonaca2012b} estimated a
distance of $26\pm4$~kpc to the stream based on fitting isochrones of
metallicity\index{metallicities} [Fe/H]=-1.0 to stars in the structure, \citet{martin13}
were able to estimate a spectroscopic metallicity\index{metallicities} of
[Fe/H]$\approx$-2.2, and revised the distance estimate to the Pisces
stream to $35\pm3$~kpc. This stream is prominent in stellar density
maps of the PAndAS\index{Pan-Andromeda Archaeological Survey (PAndAS)} survey (\citealt{martin2014}; reproduced here as
Figure~\ref{fig:pandas}; see panels 4 and 5), extending the stream
several degrees northward of the SDSS footprint. Interestingly,
\citet{martin2014} found that the stream ends abruptly, suggesting
either a physical truncation or a sharp distance gradient at the
northern end. As no progenitor has yet been identified, and as
kinematical data exist only for only one point along the stream, an
orbit has not been estimated.

{\bf Acheron\index{Acheron}, Cocytos\index{Cocytos}, Lethe\index{Lethe}, Hermus\index{Hermus}, \& Hyllus\index{Hyllus}:} These stream
candidates are long, narrow, but relatively low signal-to-noise
overdensities of low-metallicity stars in the northern footprint of
the SDSS survey. No velocity information has yet been gleaned for
these streams, but efforts are underway to associate tracers such as
blue horizontal branch stars and RR Lyrae with one or more of them.

{\bf Alpheus:}\index{Alpheus} This nearby ($< 2$ kpc) stream candidate was discovered by
combining the WISE\index{Wide-Field Infrared Survey Explorer (WISE)} All-Sky and 2MASS\index{Two Micron All Sky Survey (2MASS)} catalogs
\citep{grillmair13}. These authors suggest that, by virtue of distance
and orientation, the stream could plausibly have originated in the globular
cluster NGC 288\index{NGC 288} some $20^{\circ}$ away. Deeper surveys and velocity
information will be required to test this hypothesis.

{\bf Ophiuchus:}\index{Ophiuchus Stream} This stream is somewhat unique among our current
sample of streams in that it is apparently highly foreshortened, and
is seen nearly end-on from our vantage point. It was detected
\citep{bernard2014} as an overdensity of metal-poor stars in the
Pan-STARRS1\index{Panoramic Survey Telescope and Rapid Response System (Pan-STARRS)} 3$\pi$ survey. Though only $2.5^{\circ}$ long, Sesar et al. (2015) subsequently determined that the Ophiuchus stream is
actually $\sim1.6$~kpc in length, appearing foreshortened due to our
nearly end-on viewing angle. Their velocity\index{radial velocities} and proper motion\index{proper motions}
measurements indicate that the stream is kinematically cold and on a
fairly eccentric orbit\index{orbit fitting} with peri- and apocentric distances of 3.5 and
17.5 kpc, respectively. To explain the rather short length of the
stream given its inferred orbit, Sesar et al. suggested that the
progenitor must have suffered a substantial change in its orbit and become
disrupted only rather recently ($\sim250$~Myr ago).

\begin{figure}[!t]
\includegraphics[width=1.0\textwidth]{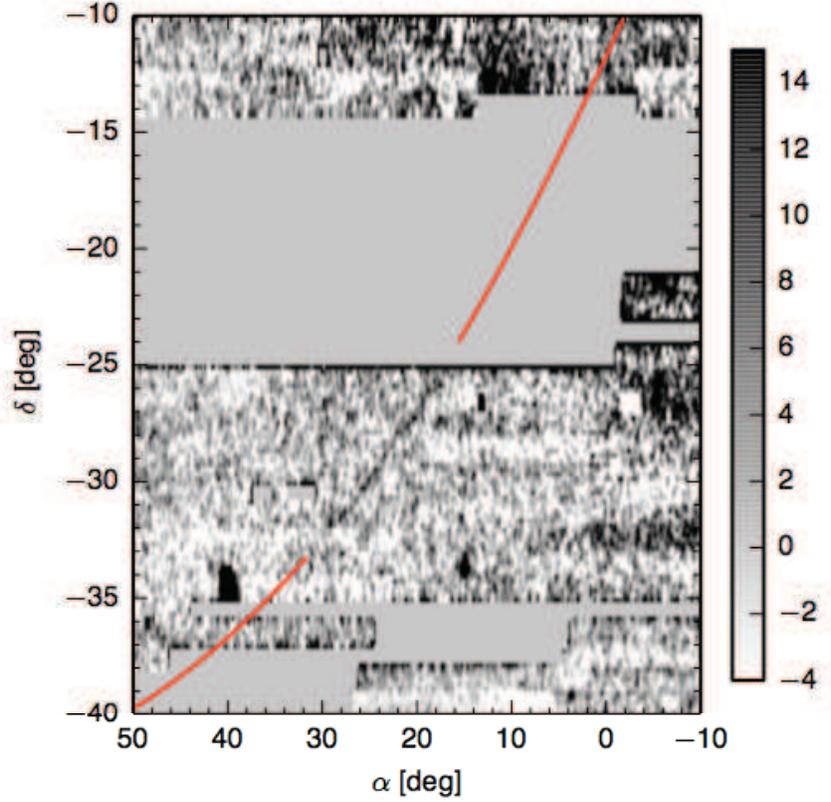}

\caption{Surface density map of the ATLAS\index{ATLAS stream} stream, using a
  filter optimized for a stellar population with [Fe/H] = -2.1, an age
  of 12.5 Gyrs, and a distance of 20 kpc. Darker shades correspond to
  higher surface densities. [Figure 1 from \citet{koposov2014}.]}

\label{fig:atlas}
\end{figure}

\begin{figure}[!t]
\includegraphics[width=1.0\textwidth]{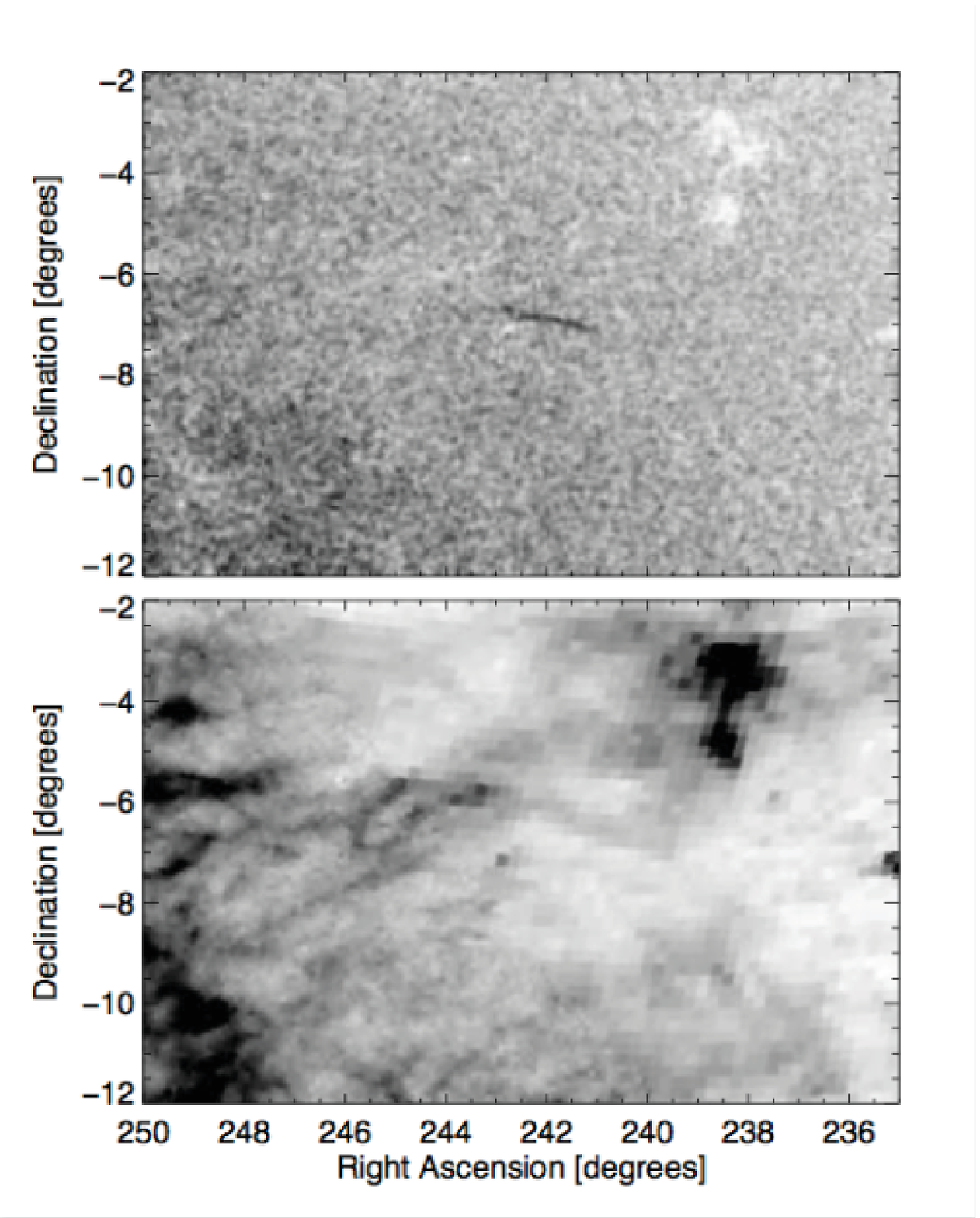}

\caption{Surface density map of the Ophiuchus stream\index{Ophiuchus Stream}, optimized for
  old and metal-poor stars at a distance of 8-12 kpc. Darker shades
  correspond to higher surface density. The lower panel shows a
  reddening map of the same region. [Adapted from Figure 1 of
  \citet{bernard2014}.]}

\label{fig:ophiuchus}
\end{figure}

% Dwarf galaxy streams:
\subsection{Streams with presumed dwarf galaxy progenitors\index{progenitors of streams!dwarf galaxies}}
\label{sec:dwarf_streams}

{\bf Orphan stream:}\index{Orphan Stream} The Orphan stream is a roughly $1-2^\circ$-wide
stellar stream spanning nearly the entire vertical (declination)
extent of the SDSS northern Galactic hemisphere footprint. This
stream, so called because its progenitor has not been identified, was
discovered at about the same time by Grillmair (2006a) and Belokurov
et al. (2007b). Because the stream has fairly low surface brightness\index{surface brightness},
determining the distance and velocity\index{radial velocities} of stream members along its
extent is difficult. However, some such measurements have been done,
and a few attempts have been made to derive orbits.  The first of
these, by Fellhauer et al. (2007a), explored the possibility that
two objects spatially coincident with the stream -- the Ursa Major II\index{Ursa Major II (UMaII)}
(UMaII) dwarf spheroidal and the HI cloud known as Complex A\index{Complex A} -- are
physically associated with the stream. They concluded that UMaII is
likely the progenitor of the Orphan stream, and predicted a strong
velocity gradient\index{velocity gradient} along the stream. In contrast, Sales et al. (2008)
modeled the disruption of a massive two-component dwarf galaxy on the
Orphan stream orbit, and found that UMaII is not likely to be the
progenitor, and indeed the progenitor\index{progenitors of streams} is likely to be almost
completely disrupted. A more extensive characterization of the stream
from SDSS was conducted by Newberg et al. (2010), who then used the
data to derive an orbit\index{orbit fitting} for the structure and produce an $N$-body
model\index{N-body simulations} of a disrupting satellite on this orbit. Their models suggest
that the SDSS debris is part of the leading tail of a disrupting
satellite on an eccentric ($e \sim 0.7$) orbit that carries the dwarf
out to Galactocentric distances of $\sim90$~kpc.

The Orphan stream\index{Orphan Stream} has been traced with RR Lyrae\index{RR Lyrae} stars to a distance of
55 kpc by Sesar et al. (2013), who also identified a strong
metallicity gradient\index{metallicity gradient} along the stream. These authors could not
identify members beyond 55 kpc, and suggest that this is either
because the leading stream actually ends at that point, or the tracers
are lost due to survey incompleteness or low metallicity. The large
metallicity\index{metallicities} spread seen among these RR Lyrae bolsters the case that
this is a dwarf galaxy remnant. The low velocity dispersion\index{velocity dispersion} and large
metallicity spread have been confirmed by \citet{casey2013}, and
high-resolution spectroscopy \citep{casey2014} shows abundances
consistent with dwarf spheroidal chemical evolution.

It is unclear whether any of the satellites (UMaII\index{Ursa Major II (UMaII)} and the Segue 1\index{Segue 1}
dwarf galaxy) or Complex A\index{Complex A} are associated with the Orphan
stream\index{Orphan Stream}. Further exploration in the southern celestial hemisphere to
identify the path of the stream, its stellar density, and possibly a
progenitor, is currently underway.  We note that this stream
is likely to be a particularly good tracer of the Galactic potential\index{Galactic potential}, since it is
well traced over a large swath (constraining its orbit quite well) and
traverses a large range of Galactic radii. The combination of the
orbit fit to the Orphan stream by Newberg et al. (2010) and distance
constraints from RR~Lyrae (Sesar et al. 2013) suggest a fairly low
mass for the Galactic halo\index{mass of Milky Way} ($\sim 3 \times 10^{11}$ M$_{\odot}$ within 60 kpc).

%Stretches across entire Dec range of SDSS, with significant distance gradient over that region. Orbit fairly well determined, but progenitor unknown. 
%Suggested that UMa (II?) and HI clouds (Complex A?) associated with it, but unlikely given orbit. 
%Passes near SegueI. 

{\bf Cetus Polar Stream:}\index{Cetus Polar Stream} This structure was identified in a study of
the Sagittarius (Sgr) stream\index{Sagittarius (Sgr) stream} (see Chapter 2) with SDSS data by Yanny
et al. (2009) as a set of BHB stars\index{blue horizontal branch (BHB) stars} with velocity\index{radial velocities} and metallicity\index{metallicities}
distinct from that of Sgr. Newberg et al. (2009) followed this up by
tracing the stream with BHB stars from SDSS DR7, finding a stream
aligned roughly along constant Galactic longitude in the south
Galactic cap. The authors showed that the ratio of blue straggler
stars (BSSs)\index{blue straggler stars (BSSs)} to BHBs is higher in Sgr than the Cetus Polar Stream\index{Cetus Polar Stream}
(CPS), and that most of the BHB stars in this region of the sky (at
distances of $\sim30$~kpc) are associated with the CPS and not
Sgr. This was confirmed by Koposov et al. (2012), who used the offset
between BHBs and BSSs to separate Sgr and the CPS in the South. The
distance and velocity trends seen in this work imply that the CPS
progenitor\index{progenitors of streams} is (or was) on an orbit that counter-rotates relative to
that of Sgr, as had also been found via orbit-fitting\index{orbit fitting} by Newberg et
al. (2009). Interestingly, Newberg et al. (2009) determined that the
massive globular cluster NGC 5824\index{NGC 5824} lies very close to their computed
orbit, at nearly the predicted distance, and with almost exactly the
predicted radial velocity. Moreover, the tidal extensions found by
\citet{grillmair95} and \citet{leon00} for NGC 5824 also lie along the
predicted orbit of the CPS.

The most extensive examination of the CPS was that of
Yam et al. (2013). This group used the additional data available in
SDSS DR8 to refine the kinematic signature and the distances to the
CPS. The newly-available contiguous sky coverage of DR8 photometry was
exploited to map the density of BHB stars\index{stellar density} along the stream and fit the
stream width. Yam et al. then used all of this new information to fit
an orbit\index{orbit fitting} to the stream, and found a low-eccentricity ($e \approx
0.2$), highly inclined ($i \approx 87^\circ$) orbit with apo- and
pericentric distances of $\sim36$ and $\sim24$~kpc from the Galactic
center. $N$-body models\index{N-body simulations} of $10^8~M_{\odot}$ satellites (assuming mass
follows light) on this orbit were found to reproduce the stream
velocities\index{radial velocities} and their dispersions, along with the width of the stream,
as a function of position. However, matching the density profile of
the stream requires a lower mass satellite of $10^6~M_{\odot}$. This
suggests that mass does not (or did not) follow light in the
progenitor of the CPS. In other words, the stream must originate from
a low-mass, dark matter-dominated satellite similar to the
``ultra-faint dwarf spheroidals'' found in SDSS. This is corroborated
by the low metallicity\index{metallicities} ([Fe/H] $\sim~-2.2$) of CPS stars, which would
place the CPS progenitor among typical metallicities of ultra-faint
dwarfs.

{\bf PAndAS MW stream:}\index{PAndAS MW Stream} This stream, seen in panels 3 and 5 of
Figure~\ref{fig:pandas}, was found by \citet{martin2014} in the PAndAS\index{Pan-Andromeda Archaeological Survey (PAndAS)}
photometric survey of the area around M31. There is a wealth of
Galactic substructure along the line of sight to M31 in this survey,
including both TriAnd~1\index{TriAnd1} and TriAnd~2\index{TriAnd2}, the Monoceros\index{Monoceros Ring} structure, and
the Pisces/Triangulum\index{Pisces/Triangulum Stream} stream. The width and velocity dispersion\index{velocity dispersion}
estimated by \citet{martin2014} suggest a dwarf galaxy progenitor\index{progenitors of streams} for
the PAndAS stream. Stellar populations suggest that this is an old,
metal-poor structure, and it follows an orbit that is roughly parallel
to the Galactic plane at a distance of $\sim17$~kpc from the Sun.

{\bf Styx:}\index{Styx} Styx is at a distance of $\approx 46$ kpc and is presumed
to be a relic of a dwarf galaxy solely by virtue of its fairly substantial
girth ($\sim 1^{\circ}$) \citep{grillmair09}. A sparse but significant
overdensity (Bo{\"o}tes~III)\index{Bo{\"o}tes III} is situated near the stream at a nearly
identical distance, and \citet{grillmair09} suggested that this
overdensity might be the remains of the stream's progenitor\index{progenitors of streams}. However,
subsequent velocity\index{radial velocities} measurements of Bo{\"o}tes~III \citep{carlin2009}
yielded $V_{hel} = 197$ km s$^{-1}$, which appears to be at odds with
the near perpendicularity of Styx to our line of sight. No velocities
have yet been measured for the stream itself.

% Clouds:
\subsection{Clouds and other diffuse stellar structures}
\label{sec:clouds}

{\bf Overdensities in Virgo:}\index{Overdensities in Virgo} The nature and number of the
substructure(s) in the Virgo constellation is uncertain. Main sequence
turnoff (MSTO) stars from SDSS in this region of the sky have been
used to show enhanced stellar densities spanning a huge sky area that
have become known as the ``Virgo Overdensity (VOD).''\index{Virgo Overdensity (VOD)} This was
initially seen in a single stripe of SDSS imaging data at a distance
of $\sim20$~kpc by \citet{newberg02}. Subsequent SDSS studies have
mapped an overdensity (relative to neighboring regions of sky) spanning
distances of at least $10 < d < 20$~kpc over $>1000$~deg$^2$
\citep{juric2008}, and subsequently over 3000~deg$^2$
\citep{bonaca2012a}. Overdensities of RR Lyrae\index{RR Lyrae} stars are seen in the
same region of sky, but in much smaller numbers. However, the much
more precise distances that can be derived for RR Lyrae have led
numerous authors to point out substructures localized in distance
rather than in a large cloud-like structure as seen in MSTO stars. The
first of these was the discovery by \citet{vivas01} of 5 RR Lyrae from
the QUEST\index{QUEST survey} survey at a distance of $\sim20$~kpc; this was followed by a
$\sim19$~kpc clump in Vivas \& Zinn (2003), and a detection at $12 < d
< 20$~kpc by Vivas \& Zinn (2006), also from QUEST. Subsequent RR
Lyrae studies from the SEKBO\index{SEKBO survey} survey by Keller et al. (2008, 2009) have
seen similar overdensities at $\sim16-20$~kpc distances. Additionally,
Keller (2010) detected an overdensity between distances of $12 < d <
18$~kpc from the Sun using subgiants from SDSS.

\begin{figure}[!t]
\includegraphics[width=1.0\textwidth]{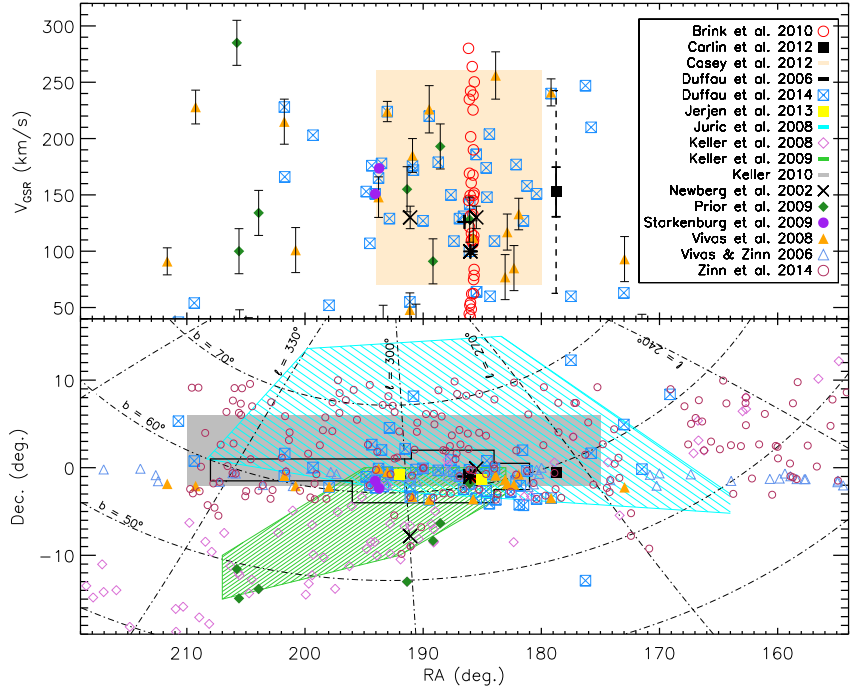}

\caption{Map\index{Overdensities in Virgo!map} of positions and velocities\index{radial velocities} of stars that have been
  associated with the Virgo substructure(s) in the literature. The
  upper panel gives GSR-frame velocities as a function of right
  ascension, while the lower panel shows RA/Dec positions on the
  sky. Sources of the data are given in the legend; those with filled
  areas on the plot are given as horizontal lines of the same color as
  the fill. While there are many ``excess'' stars in this region of
  sky at velocities higher than expected for Milky Way populations,
  there is no clear spatial overdensity or kinematically cold peak
  visible in this figure.}

\label{fig:virgo}
\end{figure}

While none of the photometric detections of Virgo substructures (see
map in Figure~\ref{fig:virgo}) are completely inconsistent with each
other, the patchiness among RR Lyrae detections suggests that there is
at the very least some density variation within a single large
structure, and perhaps a superposition of multiple structures that
makes up the overall VOD.  It is only with kinematics that this
possibility can be explored. Follow-up spectroscopy of QUEST RR Lyrae
initially showed \citep{duffau2006} that 6 stars apparently form a
cold peak at $V_{\rm gsr} \approx 100$~km~s$^{-1}$, which the authors
dubbed the ``Virgo Stellar Stream.''\index{Virgo Stellar Stream (VSS)} Subsequent studies of this
general region of sky (e.g.,
\citealt{newberg07,vivas2008,prior2009virgo,brink2010,carlin12b}) have
variously found RV peaks between 100 and 250 km~s$^{-1}$ (see
Figure~\ref{fig:virgo}), with some of these appearing to be localized
in both distance and velocity\index{radial velocities}. For example, \citet{duffau2014}
identified at least three peaks at distances ranging from $\sim10$ to
20~kpc and velocities between $\sim120-220$~km~s$^{-1}$. On the basis
of such detections, it has been argued that the substructure in Virgo
in made up of many tidal remnants overlapping in phase space. (Note,
however, that \citealt{casey2012} found velocity peaks at $V_{\rm GSR}
\sim 120$ and $\sim200-240$~km~s$^{-1}$. These authors identified the
lower-velocity peak as the VSS, and speculate that the high-velocity
feature may be related to the Sagittarius\index{Sagittarius (Sgr) stream} trailing tail.) Regardless of 
the origin, the radial velocities (see Figure~\ref{fig:virgo}) alone
make it clear that the Virgo substructure is, on average, moving too
rapidly away from the Galactic center to be a ``cloud'' of debris
piling up at orbital apocenter as discussed in Chapter~6.

An important clue to the nature of the Virgo substructures came from
the measurement of an orbit for Virgo stars in a single
pencil-beam. Initially, \citet{casetti2009} determined the orbit of a
single RR Lyrae star known to be a Virgo member. Their combined proper
motions and radial velocity yield a rather eccentric orbit that suggests this star
recently passed the pericenter of its orbit\index{orbit fitting}. This work was expanded by
\citet{carlin12b}, who measured proper motions\index{proper motions} and radial velocities\index{radial velocities}
of 17 stars consistent with VOD membership. This orbit confirms that
the Virgo substructure\index{Overdensities in Virgo} results from an object on an eccentric orbit
that recently made its closest approach. Indeed, Carlin et al. were
able to qualitatively reproduce the large spatial extent (both on the
sky and in distance) and velocity spread of the Virgo Overdensity\index{Virgo Overdensity (VOD)} by
modeling a massive dwarf spheroidal disrupting on their measured
orbit. This suggests that the entire structure may be a ``puffed-up''
dwarf galaxy that just passed its pericenter rather near the Galactic
center. However, it is unclear how to reconcile the apparent
clumpiness within the larger structure; this may be substructure
related to group infall\index{group infall} or structure within the progenitor\index{progenitors of streams} satellite.
Alternatively (as suggested by, e.g., Duffau et al. 2014), the entire
structure may be a superposition of tidal structures overlapping each
other on the sky, similar to the overlapping remnants that fell in
along ``preferred'' directions in the simulations of Helmi et
al. 2011. More observations and modeling are needed to understand the
structures in this complicated region of the sky.

{\bf TriAnd1\index{TriAnd1} and 2\index{TriAnd2}:} \citet{rochapinto04} identified an overdensity of
2MASS\index{Two Micron All Sky Survey (2MASS)} color-selected M-giant\index{M giants} candidates between roughly $100^\circ < l
< 150^\circ, -40^\circ < b < -20^\circ$. This feature, dubbed the
``Triangulum-Andromeda'' (TriAnd)\index{TriAnd1} structure, is at heliocentric
distances of $\sim18-30$~kpc.  \citet{rochapinto04} spectroscopically
confirmed that TriAnd is a kinematically cold structure, and found a
metallicity\index{metallicities} of $\langle$[Fe/H]$\rangle = -1.2$ with $\sim0.5$-dex
scatter. These authors also noted a trend of line-of-sight velocity\index{radial velocities}
with Galactic longitude that looks like an extension of the Monoceros\index{Monoceros Ring}
velocity trend; however, TriAnd was thought to be too distant to be an
extension of Monoceros. At nearly the same time, this same
substructure was seen by \citet{majewski2004triand} as a clear main
sequence in the foreground of combined CMDs from their photometric
survey of M31. These authors estimated the distance to be
$\sim16-25$~kpc from the Sun, with $\sim20\%$ higher distance at the
furthest region from the Galactic plane compared to the lowest
latitudes.  The stellar density is roughly constant over the region
surveyed, and is used to estimate a total luminosity\index{luminosity} over
$\sim1000$~deg $^2$ of $\sim5\times10^5~L_\odot$.

A much deeper and larger-area survey of the M31 outskirts presented by
\citet{martin2007} revealed the existence of a second main sequence in
the same region of the sky. This study, which covered $\sim76$~deg$^2$
between $\sim115^\circ < l < 130^\circ, -30^\circ < b < -25^\circ$,
associated the brighter of the two MSTO features with TriAnd, and the
fainter as a new structure dubbed ``TriAnd2''\index{TriAnd2} (with the original
structure called TriAnd1). These two features were found to be at
heliocentric distances of $\sim20$ and $\sim28$~kpc, and both show
fairly narrow main sequences consistent with $ \sim2$~kpc
line-of-sight depth. TriAnd1 shows stellar densities at the lowest
latitudes (i.e., nearest the plane) and lowest longitudes in the
region covered that are 2-3$\times$ higher than at the opposite side
of the survey footprint. The surface brightness\index{surface brightness} ($\Sigma_{\rm V} \sim
32~{\rm mag~arcsec} ^{-2}$) found by \citet{martin2007} is similar to
that derived by \citet{majewski2004triand} for this very diffuse
stellar substructure.

Sheffield et al. (2014) undertook an extensive spectroscopic survey of
over 200 M giant\index{M giants} stars selected to be consistent with TriAnd
membership. These authors found two distinct RGB features in 2MASS\index{Two Micron All Sky Survey (2MASS)}
that correspond to the TriAnd1\index{TriAnd1} and TriAnd2\index{TriAnd2} MSTO features from
\citet{martin2007}. Interestingly, even though the two features are
separated by more than 5 kpc in line-of-sight distance, their trends
in line-of-sight velocity\index{radial velocities} with Galactic longitude are
indistinguishable. The nearer of these features (TriAnd1) is found by
Sheffield et al. to have mean metallicity\index{metallicities} of [Fe/H] = -0.57, while
TriAnd2 has [Fe/H] = -0.64. A possible origin for the two features is
shown by these authors via an $N$-body model\index{N-body simulations} of a dwarf galaxy
disrupting in the Milky Way halo. In this scenario, the two TriAnd
features are debris that were stripped from a single satellite on
consecutive pericentric passages of its orbit. The fairly large
satellite mass required in this model is consistent with the
metal-rich (relative to most dwarf galaxies) stellar populations in
TriAnd; in order to enrich to such a level, the progenitor must have
been fairly massive.

The PAndAS\index{Pan-Andromeda Archaeological Survey (PAndAS)} deep photometric survey of the M31 vicinity has apparently
resolved some wispy substructure on smaller scales in this region of
sky. Martin et al. (2014) mapped stars from PAndAS in different
distance slices, and found a narrow stream at a distance of
$\sim$17~kpc, another wedge-shaped feature at $\sim27$~kpc that is
likely associated with TriAnd2, and wispy stellar structure throughout
the 17-kpc slice that is attributed to TriAnd1 suffusing the entire
field of view. Clearly there is a complex web of substructures
intermingling in this region of the sky. Indeed, based on positional
and kinematical similarities, \citet{deason2014} argued in a recent
analysis of spectroscopic data in this sky area that the PAndAS
stream, TriAnd overdensities, and the Segue~2\index{Segue 2} ultrafaint dwarf galaxy
are associated remnants of a group infall\index{group infall} event. This scenario posits
that the comparatively more metal-rich TriAnd represents the
group-dominant central galaxy, with Segue~2 and the PAndAS structure
remnants of TriAnd satellites. Finally, we note that Xu et al. (2015)
recently suggested that TriAnd results from the oscillation of the
midplane\index{Galactic disk!oscillation of midplane} of the disk below $b=0^\circ$, and that the apparent stellar
overdensity is part of a disk that extends out to 25 kpc or
more. There is clearly much more work needed to understand the
nature of the TriAnd features.

{\bf Hercules-Aquila Cloud:}\index{Hercules-Aquila Cloud} This structure was originally discovered
as an overdensity of MSTO stars in the SDSS DR5 database at
heliocentric distances of $\sim10-20$~kpc by Belokurov et
al. (2007b). The cloud\index{clouds} covers an enormous area of the sky, apparently
extending to at least $\pm40^\circ$ in Galactic latitude, centered on
longitude of $l\sim40^\circ$. Belokurov et al. attempted to identify
the kinematical signature of the Cloud using SDSS spectra of RGB
stars, and suggested that a peak at $V_{\rm gsr} \sim 180$~km~s$^{-1}$
represents the velocity of Hercules-Aquila stars. Due to its
similarity in the CMD to the SDSS ridgeline of the globular cluster
M92 ([Fe/H] = -2.2), the stellar population of the Hercules-Aquila was
suggested to be rather metal poor (perhaps slightly more metal-rich
than M92).

In a series of papers, Larsen et al. (2008, 2011) extensively mapped a
nearby structure between $\sim1-6$~kpc from the Sun that was dubbed
the ``Hercules Thick Disk Cloud.''\index{Hercules Thick Disk Cloud} The authors (Larsen et
al. 2011) suggest that the overdensity seen by Belokurov et al. (2007b) was
actually an artifact of the background-subtraction method used in that
work, and that the true stellar overdensity is actually much more
nearby. However, subsequent studies with unambiguous standard candles
such as RR Lyrae have confirmed that there is a structure at $d_{\rm
  hel}\sim20$~kpc. It remains unclear whether the more distant
Hercules-Aquila cloud is related to the apparent thick disk extension.

Hercules-Aquila has been extensively mapped with RR Lyrae\index{RR Lyrae} stars. Sesar
et al. (2010a) used RR Lyrae from SDSS Stripe 82 (on the
celestial equator in the south Galactic cap) to show that the
Hercules-Aquila cloud contains at least 1.6 times the stellar density
of the halo at $\sim15-25$~kpc. Photometric metallicity\index{metallicities} estimates from
this work suggest that the cloud's mean metallicity is similar to that
of the Galactic halo. Watkins et al. (2009) also studied the RR Lyrae
in SDSS Stripe 82, using the proper motions\index{proper motions} of Bramich et
al. (2008). Watkins and colleagues find 237 RR Lyrae that are likely
associated with the Hercules-Aquila cloud; these actually make up the
majority of their RR Lyrae star sample. The mean distance to the cloud
in Stripe 82 is estimated to be $\sim22$~kpc, but with dispersion
(standard deviation) of $\sim12$~kpc. Watkins et al. estimate a mean
photometric metallicity of [Fe/H] = -1.43 for the Hercules-Aquila
stars. This study additionally presents the most reliable estimate of
the velocity\index{radial velocities} of this substructure.  Stars with Hercules-Aquila cloud
metallicities separate from the thick disk at $300^\circ < RA <
320^\circ$ in velocity -- the mean velocity is centered around $V_{\rm
  gsr} \approx 25$~km~s$^{-1}$, with a long tail to lower velocities
(and blending with the disk at higher velocities).

Simion et al. (2014) mapped the Hercules-Aquila cloud using RR Lyrae\index{RR Lyrae}
from the Catalina Sky Survey\index{Catalina Sky Survey} \citep{drake2014}. After subtracting off
the contribution of the underlying halo populations, this group
concluded that the cloud is more prominent in the southern Galactic
hemisphere than in the north, peaking at a heliocentric distance of
$\sim18$~kpc. It is unclear whether the structure is truly asymmetric
about the Galactic plane, or whether higher extinction in the north
has affected the completeness of the RR Lyrae sample. The
Hercules-Aquila RR Lyrae are predominantly of Oosterhoff class I; thus
the stellar populations in the cloud are similar to those of the
Galactic halo. The luminosity\index{luminosity} of the progenitor\index{progenitors of streams} is estimated to be
between $-15 < M_{\rm V} < -9$, which would place it in the range of
the brightest ``classical'' Milky Way dwarf galaxies.

{\bf Pisces Overdensity:}\index{Pisces Overdensity} The Pisces Overdensity is one of the most
distant stellar overdensities known in the Galactic halo. This
structure was originally found in a study of Stripe~82 RR Lyrae by
Sesar et al. (2007) as a grouping of 26 stars (denoted by the authors
as ``clump J'') at a mean distance of 81~kpc from the Sun. This
structure was confirmed by Watkins et al. (2009) in a separate analysis
of RR Lyrae from SDSS Stripe~82. Watkins et al. dubbed the group of 28
stars at $\sim80$~kpc mean distance the ``Pisces Overdensity,'' and
suggested that the total stellar mass\index{stellar mass} of this substructure is
$\sim10^4 - 10^5$~M$_{\odot}$, with mean metallicity\index{metallicities} of [Fe/H]$\sim
-1.5$. Kollmeier et al. (2009) obtained follow-up spectra of a handful
of RR~Lyrae stars in Stripe~82, and found five stars tightly clumped
in velocity, confirming its identification as a coherent substructure
that the authors suggested was a (possibly disrupted) dwarf
galaxy. Additional spectra of RR~Lyrae observed by Sesar et
al. (2010b) complicated this picture by resolving the overdensity into
two peaks separated in velocity\index{radial velocities} by $\sim100$~km~s$^{-1}$. Thus the
Pisces Overdensity may actually consist of more than one tidal debris
feature overlapping in space. Sharma et al. (2010) subsequently found
a much larger extended structure (dubbed ``A16'') in the same region
of sky using 2MASS-selected\index{Two Micron All Sky Survey (2MASS)} M-giant\index{M giants} candidates. In this work, the
candidates are at distances of $\sim100$~kpc. The large spatial extent
(including another possibly associated structure, ``A14,'' at similar
distances about 30 degrees away) and somewhat metal-rich (as evidenced
by the presence of M-giants) nature of this structure was suggested to
indicate an unbound satellite.

While the origin and nature of the Pisces Overdensity remains unclear,
it is evident that there are tidal debris structures at distances of
more than 80~kpc in the halo. Further kinematical and chemical
analysis will be essential to relate the Pisces substructure to other
known satellites or structures in the halo.

{\bf Perseus Cloud:}\index{Perseus Cloud} Another poorly studied apparent substructure is
the Perseus Cloud, originally isolated among 2MASS-selected M-giant\index{M giants}
candidates by Rocha-Pinto et al. (2004). Though it appears to be an
extension of the TriAnd structure on the sky, and is at a similar
distance, Rocha-Pinto et al. argue that their spectroscopic velocities\index{radial velocities}
show the two structures to be unrelated. The nature of this stellar
structure remains undetermined.

{\bf Canis Major\index{Canis Major} and Argo\index{Argo}:} There are other purported stellar
overdensities in the Milky Way that we have not discussed here because
their nature (as an actual overdensity and/or as a tidal debris
structure) is still undetermined. These include the Canis Major
overdensity, which was originally seen as a low-latitude excess of
2MASS-selected\index{Two Micron All Sky Survey (2MASS)} M~giants\index{M giants} by \citet{martin2004a}. This feature has been
suggested to be a recently accreted dwarf galaxy \citep{martin2004a, martin2004b,
 bellazzini2006}, and perhaps even the progenitor of the Monoceros
Ring\index{Monoceros Ring} (e.g., \citealt{penarrubia05}). Others (including Momany et
al. 2006, Moitinho et al. 2006) have attributed this apparent stellar
overdensity to the warp and flare\index{Galactic disk!warp and flare} of the Galactic disk. The Argo
stellar overdensity (Rocha-Pinto et al. 2006) is near the Canis
Major feature in 2MASS M~giants, and is also poorly studied. As the 
focus of this review is on known tidal substructures, and the nature
of these structures is still under debate, we will not discuss them
further. Some discussion of Canis Major and its possible relation
to the low-latitude Monoceros structure can be found in Chapter~3 of
this volume.

%Controversial -- may be an actual overdensity, or may be a reddening window (it's at low latitude). Has been suggested as progenitor of Monoceros.

%Argo: Rocha-Pinto et al. 2006\\
%& & $-71^{\circ} < \delta < 0^{\circ}$ \\

%Canis Major   & dG?& $95^{\circ} < $ R.A. $< 120^{\circ}$ & 7.2 & 109
%& -0.5 & Martin et al 2004a,b, Bellazzini et al. 2004, 2006, \\
%& & $-45^{\circ} < \delta < -10^{\circ}$ & & & & Dinescu et al. 2005 \\

%Large cloud-like overdensities thought to be pile-ups of tidal debris at apocenter. 

\section{Future Discoveries}

Given that the currently available, large-area digital sky surveys do not
cover the entire sky, we expect that a significant number of strong
streams remain to be discovered within 50 kpc. The Pan-STARRS\index{Panoramic Survey Telescope and Rapid Response System (Pan-STARRS)} survey
(Kaiser et al. 2002), which covers considerably more sky area than the
SDSS\index{Sloan Digital Sky Survey (SDSS)}, will presumably yield additional streams, and extend at least
some of the streams whose known extent is currently limited by the
SDSS footprint. In the southern hemisphere, the ATLAS\index{ATLAS} (Shanks et
al. 2013), Skymapper\index{Skymapper} (Keller et al. 2007), and Dark Energy Surveys\index{Dark Energy Survey (DES)} (Rossetto et al. 2011) remain to be completed. Covering regions of the
sky that have been largely unexplored to date, they will almost
certainly yield a number of new streams, in addition to extending some
of the streams and clouds in Tables 1 and 2. We note that extending known
streams is at least as important as finding new ones, as longer
streams both reduce the uncertainty in orbit shape and tighten the
possible constraints on the Galactic potential.

Further progress in finding cold or tenuous streams in the SDSS and
other surveys will almost certainly result from simultaneously combining photometric
filtering\index{matched filter} with velocity\index{radial velocities} and proper motion\index{proper motions} measurements. Spectra are
now available for $\sim 7 \times 10^5$ stars in the SDSS DR10. Newberg
et al. (2009) combined photometric measurements with velocities and
metallicities\index{metallicities} to detect the Cetus Polar Stream, a structure which had
eluded prior discovery through photometric means alone.  Several
spectroscopic surveys either planned or in progress (e.g., LAMOST\index{Large Sky Area Multi-Object Fiber Spectroscopic Telescope (LAMOST)}) will
almost certainly make contributions in this area, particularly in
detecting streams that are highly inclined to our line of sight
(e.g., ECHOS\index{Elements of Cold HalO Substructure (ECHOS)}, Schlaufman et al. 2009).

The Large Synoptic Survey, expected to begin early in the next decade,
will greatly extend the reach of current photometric (e.g., matched
filter\index{matched filter}) methods of stream detection. Even a single pass over the
visible sky will be significantly deeper than the SDSS, and the
end-of-survey photometric depth is expected to reach $r = 27.5$
(Grillmair \& Sarajedini 2009). On the other hand, background galaxies
will vastly outnumber stars at this depth, and significant
improvements will need to made in the area of star-galaxy separation
if we are to take full advantage of the data. If this can be achieved,
then applying matched-filter techniques to upper main sequence stars
should enable us to photometrically detect streams out to as much as
500 kpc. RR Lyrae\index{RR Lyrae} will enable us to probe even more deeply into the
halo (e.g. Sesar et al. 2013); even with fairly conservative
single-pass detection limits, LSST should enable us to detect RR Lyrae
out to more than 700 kpc.

Distance estimates are currently the largest source of uncertainty in
the use of tidal streams as probes of the potential. The Gaia\index{Gaia} survey
will clearly have an enormous impact here, providing {\it
  trigonometric} distances for giant branch stars out to tens of
kpc. In addition, a number of synoptic surveys have detected many
thousands of RR Lyrae, some of which have been associated with known
streams (e.g. Sesar et al. 2013), and whose distances can now be
measured to $\simeq 2$\% using infrared photometry. On the other hand, the
coldest streams may have very few giant branch stars or RR
Lyrae. Distances for these have often been estimated photometrically, 
using filter matching or upper main sequence fitting. However, the ages
and metallicities\index{metallicities} of the stream stars are only poorly known, and the
distance uncertainties due to inappropriate filter templates are
correspondingly large. Eyre \& Binney (2009) and Eyre (2010) have
described and demonstrated an alternative method of finding distances
by using proper motions\index{proper motions} to measure ``Galactic parallax.''\index{Galactic parallax} This
technique relies on the premise that any net proper motion component
of stars perpendicular to a stream must be a consequence of the sun's
reflex motion. The accuracy of the method is currently on par with
photometric distances, limited primarily by the quality of the
available proper motions. The Gaia survey should enable very significant
improvements in the application of this technique.

Proper motions\index{proper motions} can also be used to verify and even detect streams, and
both Gaia\index{Gaia} and LSST\index{Large Synoptic Survey Telescope (LSST)} are expected to make significant contributions to
the field.  With expected end-of-mission proper motion accuracies of
$\le 200 \mu$as to 20th magnitude, the Gaia\index{Gaia} survey will enable us to
push well down the stream mass function out to perhaps tens of
kpc. Beyond that, the end-of-survey proper motion uncertainties for
LSST\index{Large Synoptic Survey Telescope (LSST)} will be on the order of 100 km s$^{-1}$ at 100 kpc. With suitable
averaging and the application of Bayesian techniques, this should be
sufficient to detect the remote and particularly high-value, extended
tidal tails of many of the dwarf galaxies.

\begin{acknowledgement}
JLC gratefully acknowledges support from the NSF under grants AST 09-37523 and AST 14-09421.
\end{acknowledgement}
%
%\section*{Appendix}
%\addcontentsline{toc}{section}{Appendix}
%
%
%When placed at the end of a chapter or contribution (as opposed %to at the end of the book), the numbering of tables, figures, and equations in the appendix section continues on from that in %the main text. Hence please \textit{do not} use the \verb|appendix| command when writing an appendix at the end of your chapter or contribution. If there is only one the appendix %is designated ``Appendix'', or ``Appendix 1'', or ``Appendix 2'', etc. if there is more than one.

%\input{bibliography_all}

%\backmatter
%\printindex
%
\end{document}